\documentclass[]{mn2e}
\usepackage{epsfig,amsmath}
\def\gsim{\vcenter{\hbox{$>$}\offinterlineskip\hbox{$\sim$}}}
\def\lsim{\vcenter{\hbox{$<$}\offinterlineskip\hbox{$\sim$}}}
\title[H\,{\sc i} near Pal\,4]{A peculiar H\,{\sc i} cloud near the distant
globular cluster Pal\,4}
\author[Jacco Th. van Loon et al.]{Jacco Th. van Loon$^{1}$, Sne\v{z}ana
Stanimirovi\'c$^{2}$, Mary Putman$^{3}$, Joshua E.G. Peek$^{4}$,
\newauthor Steven J. Gibson$^{5}$, Kevin A. Douglas$^{6}$, and Eric J.
Korpela$^{7}$\\
$^{1}$Astrophysics Group, Lennard-Jones Laboratories, Keele University,
      Staffordshire ST5 5BG, United Kingdom\\
$^{2}$Department of Astronomy, University of Wisconsin, Madison, WI 53706,
      USA\\
$^{3}$Department of Astronomy, Columbia University, New York, NY 10027, USA\\
$^{4}$Department of Astronomy, UC Berkeley, 601 Campbell Hall, Berkeley, CA
      94720, USA\\
$^{5}$Department of Physics \& Astronomy, Western Kentucky University, Bowling
      Green, KY 42101, USA\\
$^{6}$Astrophysics Group, School of Physics, University of Exeter, Stocker
      Road, Exeter EX4 4QL, United Kingdom\\
$^{7}$Space Sciences Laboratory, UC Berkeley, 601 Campbell Hall, Berkeley, CA
      94720, USA}
\date{Resubmitted 5 March 2009}
\pagerange{\pageref{firstpage}--\pageref{lastpage}}
\pubyear{2009}
\begin{document}
\maketitle
\label{firstpage}
\begin{abstract}
We present 21-cm observations of four Galactic globular clusters, as part of
the on-going GALFA-H\,{\sc i} Survey at Arecibo. We discovered a peculiar
H\,{\sc i} cloud in the vicinity of the distant (109 kpc) cluster Pal\,4, and
discuss its properties and likelihood of association with the cluster. We
conclude that an association of the H\,{\sc i} cloud and Pal\,4 is possible,
but that a chance coincidence between Pal\,4 and a nearby compact
high-velocity cloud cannot be ruled out altogether. New, more stringent upper
limits were derived for the other three clusters: M\,3, NGC\,5466, and
Pal\,13. We briefly discuss the fate of globular cluster gas and the
interaction of compact clouds with the Galactic Halo gas.
\end{abstract}
\begin{keywords}
ISM: clouds ---
globular clusters: individual: Pal\,4, M\,3, NGC\,5466, Pal\,13 ---
Galaxy: halo ---
galaxies: dwarf ---
dark matter ---
radio lines: ISM
\end{keywords}

%=========================================================================== 1
\section{Introduction}

The Galactic globular cluster system comprises about 150 clusters that are
distributed throughout the Halo (Harris 1996). Though mostly concentrated
towards the Galactic Bulge, five clusters (AM\,1, Eridanus, NGC\,2419, Pal\,3,
and Pal\,4) are known at Galacto-centric distances around $\sim100$ kpc ---
twice as far as the Magellanic Clouds! They are excellent probes of the
gravitational potential of the Halo. They also probe the tenuous, hot
($\sim10^6$ K) gas permeating the Galactic Halo (Spitzer 1956; Sembach et al.\
2003; Bregman 2007), in particular through ram-pressure due to matter shed by
the red giants in the clusters. These are evolved low-mass ($\sim0.8$
M$_\odot$) stars that will lose $\sim0.3$ M$_\odot$ of material to the
cluster's ISM before becoming $\sim0.5$ M$_\odot$ white dwarfs (e.g., Rood
1973; Tayler \& Wood 1975; van Loon, Boyer \& McDonald 2008; McDonald et al.\
2009). The main constituent of red giant winds is neutral hydrogen; with
typical wind speeds $\sim10$ km s$^{-1}$ being slower than the escape velocity
of a typical globular cluster (McDonald \& van Loon 2007), there is a fair
chance to find diffuse neutral hydrogen gas within globular clusters.

The globular cluster gas (GCG) has proven to be elusive, the only convincing
detections being infrared emission from dust and 21-cm emission of atomic
hydrogen in M\,15 (Evans et al.\ 2003; Boyer et al.\ 2006; van Loon et al.\
2006) and plasma in 47\,Tuc deduced from pulsar timing experiments (Freire et
al.\ 2001). The associations of H\,{\sc i} clouds with NGC\,2808 (Faulkner et
al.\ 1991) and M\,56 (Birkinshaw, Ho \& Baud 1983) are uncertain due to their
low Galactic latitudes. The paucity of observed GCG suggest that gas may be
removed from the cluster on timescales as short as $<10^6$ yr (McDonald et
al.\ 2009). X-ray emission presumed to arise from the bow-shock interface
between the GCG and the Halo gas was recently detected ahead of several
globular clusters (Okada et al.\ 2007), suggesting that these interactions may
be commonplace (cf.\ Faulkner \& Smith 1991; Krockenberger \& Grindlay 1995).
This confirms the ram-pressure stripping mechanism proposed by Frank \& Gisler
(1976), which Priestley, Salaris \& Ruffert (2008) show lead to a retention of
less than a solar mass of gas.

If the gas leaving globular clusters is by and large warm neutral or weakly
ionized gas, and if it retains some degree of compactness, then there might be
a possibility to observe these clouds externally to clusters, perhaps
ressembling compact high-velocity clouds (compact HVCs), which are found
throughout the Galactic Halo (Muller et al.\ 1963; Wakker \& van Woerden 1997;
Braun \& Burton 1999). With previous searches for GCG having concentrated on
the cores of a small, biased sample of (generally nearby) globular clusters, a
sensitive, uniform large-areal H\,{\sc i} survey is called for to detect GCG
internal or external to globular clusters.

We here report the first results of an on-going project that forms part of the
Galactic Arecibo L-band Feed Array (GALFA) H\,{\sc i} Survey presently being
conducted at the Arecibo radio observatory. It has the following two
objectives: (i) to search for hydrogen gas originating within Galactic
globular clusters, and (ii) to search for evidence of interaction between
globular clusters and the interstellar media of the Galactic Halo and Disc. We
concentrate on Pal\,4, in which vicinity we discovered an unusual H\,{\sc i}
cloud. As one of the most distant globular clusters known, its study may
reveal properties of what is essentially {\it Terra Incognita} in the outer
Galactic Halo. Non-detections are reported for three further clusters, viz.\
M\,3, NGC\,5466, and Pal\,13, providing the most stringent limits yet on the
neutral hydrogen content in these clusters.

%=========================================================================== 2
\section{Arecibo 21 cm observations}

The data were collected at the Arecibo radio observatory\footnote{The Arecibo
Observatory is part of the National Astronomy and Ionosophere Center, which is
operated by Cornell University under a cooperative agreement with the National
Science Foundation}, Puerto Rico, over the course of 2005--2007, as part of
the Turn On GALFA Survey (TOGS). This survey runs commensally with
extra-galactic drift-scan surveys as part of the Galactic Arecibo L-band Feed
Array (GALFA) H\,{\sc i} Survey (Stanimirovi\'c et al.\ 2006), but crossing
scans from other GALFA projects have been added whenever possible to improve
calibration and remove scanning artifacts. The data have a beam and channel
width of $3.5^\prime$ and 0.2 km s$^{-1}$, respectively; we worked with
datacubes sampled at $1^\prime$ and 0.736 km s$^{-1}$, respectively, covering
$\pm74$ km s$^{-1}$ around the velocity of the globular cluster target.
Technical information about the observing strategy and basic data processing
can be found in Peek \& Heiles (2009).

%
% TABLE 1
%
\begin{table*}
\caption[]{Globular clusters observed as part of the TOGS survey as of early
2008. Distances are given with respect to the Sun ($d_\odot$), Galactic Centre
($d_{\rm GC}$), and the Galactic plane ($h$), and $R_{\rm t}$ is the tidal
radius (the half-light radii are comparable to, or smaller than, the Arecibo
beam). Data are taken from the Harris (1996) catalogue, except the value for
$v_{\rm LSR}$ of Pal\,13 which is from C\^ot\'e et al.\ (2002), and the
cluster masses ($M$) which we derived from $M_{\rm V}$ employing the scaling
relationship of Mandushev, Spassova \& Staneva 1991). The 3-$\sigma$ noise
levels are given per beam and per (projected) tidal sphere ($V_{\rm t}$), in
M$_\odot$ of atomic hydrogen. These can be considered upper limits to the
possible presence of GCG, and in the case of Pal\,4 they do {\it not} include
the nearby HVC.}
\begin{tabular}{lcccccccccccc}
\hline\hline
Cluster             &
RA \& Dec           &
                    &
$v_{\rm LSR}$       &
$d_\odot$           &
$d_{\rm GC}$        &
$h$                 &
$M_{\rm V}$         &
$\log M$            &
[Fe/H]              &
$R_{\rm t}$         &
\multicolumn{2}{c}{H\,{\sc i} 3-$\sigma$ limit} \\
                    &
(J2000)             &
                    &
(km s$^{-1}$)       &
(kpc\rlap)          &
(kpc)               &
\llap{(}kpc\rlap{)} &
\llap{(}mag\rlap{)} &
(M$_\odot$)         &
                    &
($^\prime$)         &
\llap{(}M$_\odot$/beam\rlap{)} &
\llap{(}M$_\odot/V_{\rm t}$)   \\
\hline
Pal\,4               &
11 29 16.8 +28 58 2\rlap{5} &
                     &
$76.7\pm2.1$         &
\llap{1}09.2         &
\llap{1}11.8         &
103.7                &
$-$6.02              &
4.4                  &
\llap{$-$}1.48       &
3.\rlap{3}           &
\llap{3}0            &
\llap{5}1            \\
M\,3                 &
13 42 11.2 +28 22 3\rlap{2} &
                     &
\llap{$-1$}$37.9\pm0.4$ &
10.4                 &
12.2                 &
10.2                 &
$-$8.93              &
5.7                  &
\llap{$-$}1.57       &
\llap{3}8.\rlap{2}   &
0\rlap{.5}           &
\llap{1}0            \\
NGC\,546\rlap{6}     &
14 05 27.3 +28 32 0\rlap{4} &
                     &
\llap{1}$19.7\pm0.3$ &
17.0                 &
17.2                 &
16.3                 &
$-$7.11              &
4.9                  &
\llap{$-$}2.22       &
\llap{3}4.\rlap{2}   &
1\rlap{.6}           &
\llap{2}8            \\
Pal\,13              &
23 06 44.4 +12 46 1\rlap{9} &
                     &
$30.4\pm0.5$         &
26.9                 &
27.8                 &
\llap{$-$}18.3       &
$-$3.51              &
3.2                  &
\llap{$-$}1.65       &
2.\rlap{2}           &
4\rlap{.3}           &
6                    \\
\hline
\end{tabular}
\end{table*}

To optimise the sensitivity of the datacube to small H\,{\sc i} clouds, and to
remove artifacts resulting from the scanning technique, we performed the
following operations (in {\sc idl}): (i) apply a velocity median-filter with a
running 5-channel boxcar; (ii) correct for row offsets (horizontal striping)
by subtracting the median of all columns; (iii) correct for the baselines in
the spectra (continuum subtraction) by subtracting the median value in the
spectrum; (iv) rebin the velocity channels by a factor 4; (v) apply a spatial
average-filter with a running $3\times3$ pixels boxcar; and finally (vi) rebin
spatially by a factor $2\times2$. The resulting root-mean-square brightness
temperature noise in the smoothed/filtered maps is $\sigma(T_{\rm B})\sim15$
mK.

%
% FIGURE 1
%
\begin{figure}
\centerline{\psfig{figure=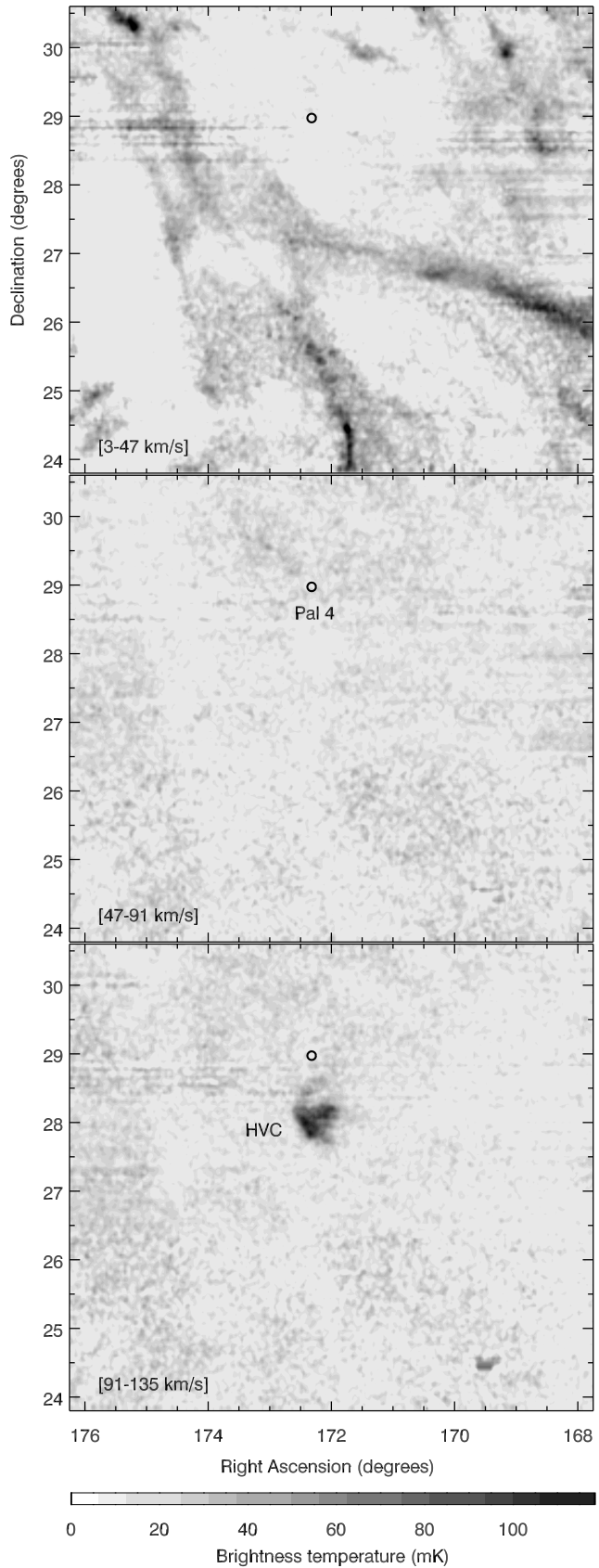,width=84mm}}
\caption[]{The 21-cm images around Pal\,4 at $3^\prime$ resolution, in
velocity ranges near H\,{\sc i} emission from Galactic cirrus (top), around
the systemic velocity of Pal\,4 (indicated with a circle; middle), and
comprising the high-velocity cloud (HVC; bottom).}
\end{figure}

%=========================================================================== 3
\section{Globular cluster targets}

The targets are all globular clusters that were covered by the TOGS data as of
early 2008 (Table 1). Although just four for the moment, these are interesting
clusters: Pal\,4 is the second-most distant cluster (after AM\,1), M\,3 is a
rather massive cluster whilst Pal\,13 is one of the dimmest clusters ---
believed to be near dissolution unless bound by non-luminous matter (Siegel et
al.\ 2001; C\^ot\'e et al.\ 2002). NGC\,5466 has a metallicity at the lower
end of the Galactic globular clusters.

Warm GCG is expected to be in equilibrium with the gravitational potential,
although a cool GCG would sink to the cluster centre (van Loon et al.\ 2006).
One might think that M\,3 presents the highest potential for detecting GCG
because it may have been able to retain it within its deep gravitational
potential, or alternatively Pal\,4 because it may have been less perturbed and
hence been able to accumulate more GCG. Both M\,3 and NGC\,5466 have a large
projected tidal radius, possibly resulting in GCG of low surface brightness,
whilst the small projected tidal radii of Pal\,4 and Pal\,13 could cause their
GCG to under-fill the Arecibo beam thus diluting the signal. The low
metallicity of NGC\,5466 does not preclude the presence of H\,{\sc i}, given
that H\,{\sc i} emission was detected from the equally metal-poor cluster
M\,15.

%=========================================================================== 4
\section{A peculiar H\,{\sc i} cloud near Pal\,4}

%
% FIGURE 2
%
\begin{figure}
\centerline{\psfig{figure=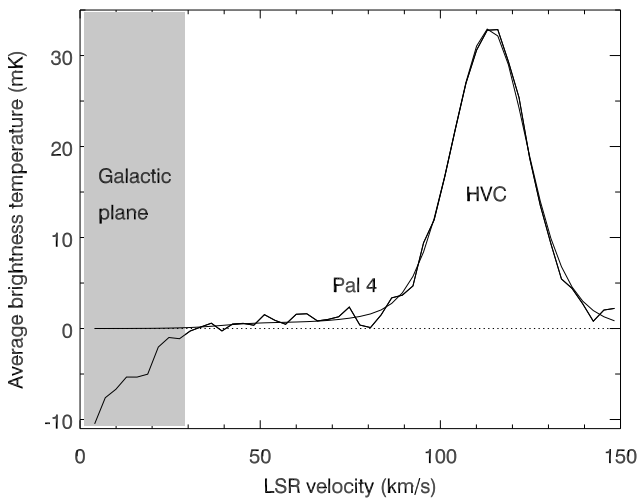,width=84mm}}
\caption[]{Line profile of the 21 cm emission in the region of Pal\,4 and the
nearby high-velocity cloud, averaged over a $1.15^\circ\times1.25^\circ$ area.
The profile dips below zero as a result of bright Galactic plane emission
affecting the striping and baseline corrections.}
\end{figure}

The Pal\,4 data are intriguing: a bright cloud is found a degree to the South
of the cluster (Fig.\ 1), offset in velocity by $\sim40$ km s$^{-1}$ (Fig.\
2). The surrounding area and velocity range are otherwise rather clean, with
filamentary Galactic plane emission constrained to $v_{\rm LSR}<30$ km
s$^{-1}$ (the little smudge at $RA=169.5^\circ$, $Dec=24.5^\circ$ in the HVC
panel of Fig.\ 1 is an artefact due to missing data). The velocity difference
with respect to the maximum extension of the Galactic plane emission is $>50$
km s$^{-1}$, classifying the cloud as a {\it bona fide} high-velocity cloud
(HVC; Wakker 1991), and as it subtends $<2^\circ$ on the sky it belongs to the
subclass of {\it compact} HVCs (Braun \& Burton 1999). The weak tail of
emission closer to Pal\,4 has a velocity closer to that of the cluster.

This HVC is not listed in the catalogue of compact HVCs of de Heij, Braun \&
Burton (2002a). Birkinshaw et al.\ (1983) observed Pal\,4 with Arecibo at 21
cm, but they did not detect the HVC as they only pointed at the cluster.

The HVC was analysed by fitting Gaussian functions to all the spectra in the
datacube as processed until step (iv) described in Section 2. The fits were
constrained within reason, and were successful even at a low signal level.
There was no need to invoke multiple components. The fits yield maps of the
peak brightness temperature, line width, and velocity centroid (Fig.\ 3). The
line width is converted into an apparent gas temperature according to
\begin{equation}
\sigma_{\rm v} = (kT/m)^{0.5},
\end{equation}
where $T$ is the temperature, $m$ the mass of the hydrogen atom, and $k$ the
constant of Stefan-Boltzmann. Apart from thermal motions this temperature may
include bulk motions due to turbulence or other kinematic structures.

%
% FIGURE 3
%
\begin{figure*}
\centerline{\psfig{figure=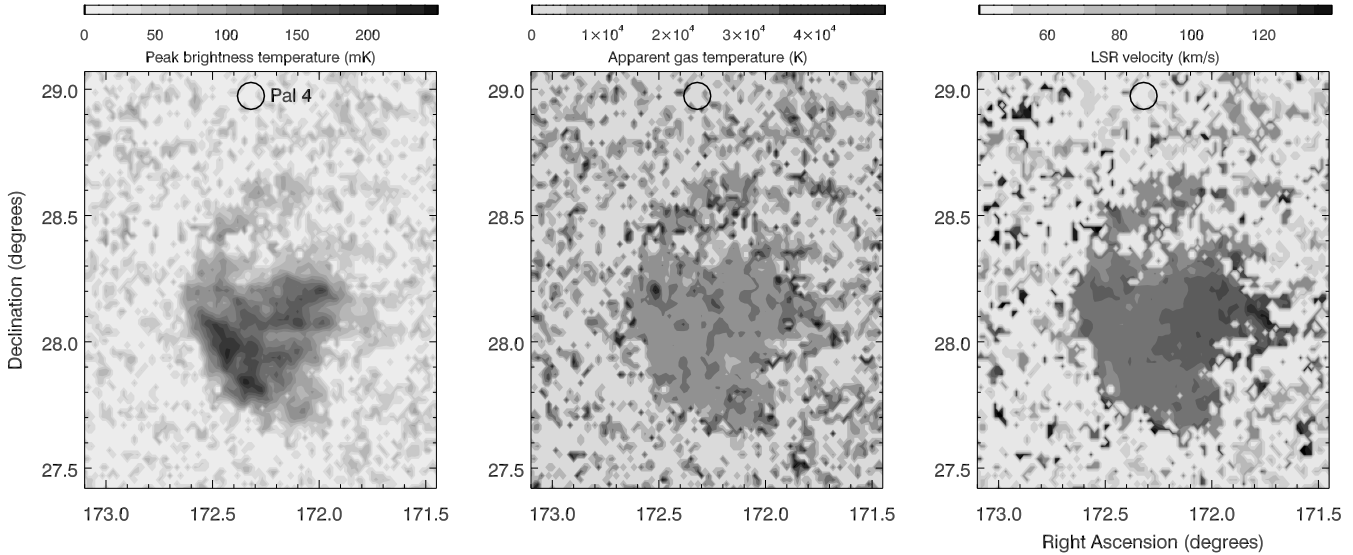,width=178mm}}
\caption[]{Peak brightness temperature (left), apparent gas temperature
(middle --- see text for a discussion of its meaning and caveats), and
velocity centroid (right) maps constructed by Gausssian fits to the datacube.}
\end{figure*}

As in Fig.\ 1, the circle in Fig.\ 3 indicates the position of Pal\,4, with
the circle diameter approximately equal to the tidal sphere; the HVC is
located outside the gravitational dominance of Pal\,4. The brightest portion
of the HVC spans $\sim$half a degree in diameter, corresponding to $\sim1$ kpc
if at the distance of Pal\,4. It has a notably bright South-Eastern rim, and
otherwise displays small-scale structure down to the angular resolution limit
of Arecibo --- a physical scale of $\sim100$ pc at the distance of Pal\,4. The
HVC appears fairly isothermal, and with $T_{\rm Cloud}\sim10,000\pm5,000$ K
typical of a warm diffuse medium. However, this is an upper limit as it
includes any non-thermal contributions to the line width (e.g., turbulence).
Hence, colder gas may still be present.

The H\,{\sc i} mass of the HVC can be estimated from
\begin{equation}
M_{\rm H}({\rm M}_\odot)=0.024\ d^2 (\delta\omega/\Delta\Omega)^2 \iint T_{\rm
B}(v)\,{\rm d}v\,{\rm d}A
\end{equation}
(cf.\ Braun \& Burton 2000), where $d$ is the distance (in kpc),
$\delta\omega$ and $\Delta\Omega$ are the pixel size and beam width, and the
integration over $A$ is simply obtained by summing the pixel values over the
area covered by the H\,{\sc i} emission, with brightness temperature $T_{\rm
B}(v)$ in K and velocities $v$ in km s$^{-1}$. In reality the beam is
elliptical, but the resulting reduction in the mass estimate is small for a
declination of $+28^\circ$ (Heiles et al.\ 2001). Integrating over a 44 km
s$^{-1}$ interval centred on 113 km s$^{-1}$, one obtains a mass of $M_{\rm
H}=8\times10^4 (d/109\ {\rm kpc})^2$ M$_\odot$, or $M>10^5 (d/109\ {\rm
kpc})^2$ M$_\odot$ if accounting for helium and perhaps additional molecular
gas or hot plasma.

Although Miville-Desch\^enes et al.\ (2005) detected dust in the high-velocity
Complex C, most HVCs lack dust (Peek et al.\ 2009) and especially the
metal-poor ones have little H$_2$ (Richter et al.\ 2001). We examined the
DIRBE/IRAS dust maps created by Schlegel, Finkbeiner \& Davis (1998), but
could not find any evidence for the presence of dust in the HVC near Pal\,4.
This is not surprising, especially if the HVC originates within this
metal-poor globular cluster.

The HVC appears to display a velocity gradient across its surface, differing
by $\Delta v\sim10$ km s$^{-1}$ between its Eastern and Western extremities.
We dismiss the possibility that this is a result of perspective motion. In
the small angle limit --- which can be shown {\it a posteriori} to be valid
--- the absolute space motion of the HVC would have to be
$v\approx\Delta v/\delta$, where $\delta\sim0.6^\circ$ is the angle on the sky
between the Eastern and Western extremities, i.e.\ $\sim1000$ km s$^{-1}$.
This would make it a {\it hyper}-velocity cloud, unbound to the Local Group.

If the velocity gradient were due to gravitational force, then a mass of
$M_{\rm rot}=(\Delta v/2)^2 r/G\sim3\times10^6\ (d/109\ {\rm kpc})$ M$_\odot$
is inferred, for $r\sim0.5\ (d/109\ {\rm kpc})$ kpc. It is interesting to note
that an essentially identical result is obtained if the virial theorem is
applied instead, $M_{\rm vir}= (5/3) (\sigma_{\rm v})^2 r/G$, to relate the
measured kinetic energy at {\it small} scales, $\sigma_{\rm v}\sim10$ km
s$^{-1}$ (Fig.\ 3, middle), to the gravitational potential. Including the {\it
global} kinetic structure (i.e.\ the velocity gradient) in this equation by
replacing $\sigma_{\rm v}$ by the FWHM of the spatially-integrated line
profile (Fig.\ 2), $\sim20$--25 km s$^{-1}$, would yield an even larger mass
by a factor $\sim20$; but this would be wrong as such kinetic structure is
incompatible with it being a virialized (dynamically relaxed) system.

If the cloud is gravitationally confined, then the total-to-baryonic mass
ratio, $M_{\rm tot}/M_{\rm bar}\ \lsim\ 30$ $(109\ {\rm kpc}/d)$, depending on
the molecular and ionized content (Pal\,4 itself is outside the dynamical
region considered here). If all baryonic mass were atomic then $M_{\rm
bar}\sim M_{\rm H+He}$. For the atomic mass detected via H\,{\sc i} to account
for all of the dynamical mass ($M_{\rm tot}$), the distance would have to be
$d\sim3$ Mpc. At such large distance the cloud would be $\sim30$ kpc in
diameter, i.e.\ the size of a typical spiral galaxy. If the velocity gradient
is indeed due to rotation under the influence of gravity, then this ratio {\it
increases} if the cloud is nearer to us.

But velocity gradients of this size are often seen in HVCs (e.g., Westmeier,
Br\"uns \& Kerp 2005). The dynamical mass estimate may therefore not be valid
and the cloud mass may be similar to the estimate based purely on H\,{\sc i}.
In some cases velocity gradients result from a superposition of several
H\,{\sc i} clouds (e.g., Stanimirovi\'c et al.\ 2002). Yet the HVC near
Pal\,4 is not part of an extensive complex of H\,{\sc i} clouds such as make
up the Magellanic Stream, so one then wonders about the cause for --- and
sustainability of --- the kinematic substructure within this isolated cloud.

%=========================================================================== 5
\section{Upper limits on neutral hydrogen within M\,3, NGC\,5466, Pal\,4 and
Pal\,13}

%
% FIGURE 4
%
\begin{figure*}
\centerline{\psfig{figure=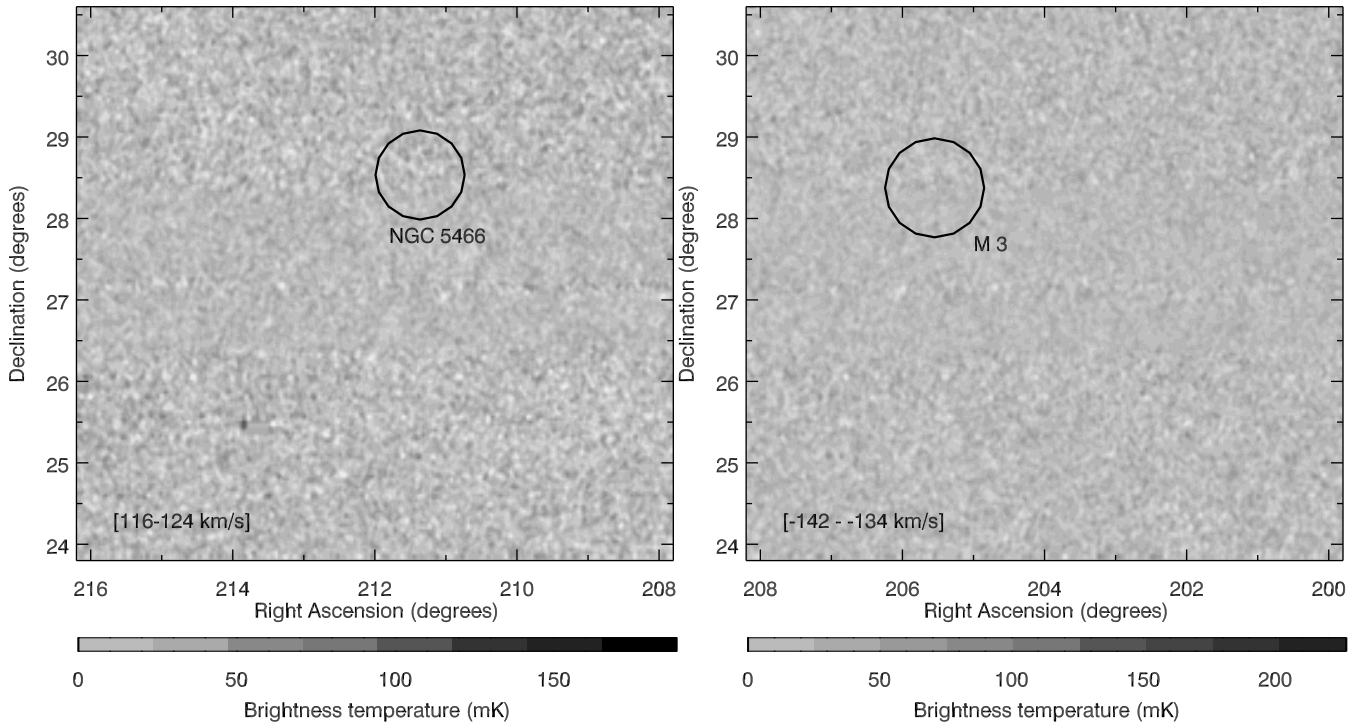,width=178mm}}
\caption[]{The 21-cm images around NGC\,5466 (left) and M\,3 (right), in a 9
km s$^{-1}$ velocity range around the H\,{\sc i} emission at each of their
systemic velocities. The size of the circle represents the extent of the tidal
sphere in each case.}
\end{figure*}

No gas is found within any of the globular clusters, M\,3, NGC\,5466, Pal\,4
or Pal\,13. We have estimated 3-$\sigma$ noise levels, expressed in M$_\odot$
of atomic hydrogen mass contained within an Arecibo beam as well as within the
projected tidal sphere of the cluster, valid for H\,{\sc i} emission
originating at the cluster distance (Table 1).

The M\,3 and NGC\,5466 data show no H\,{\sc i} emission at all (Fig.\ 4).
These clusters have tidal radii in excess of half a degree, which results in
relatively high upper limits to the hydrogen mass if spread out over the tidal
sphere. A tidal tail was discovered associated with NGC\,5466 (Belokurov et
al.\ 2006), and the gas may have been susceptible to tidal stripping as well.
The stellar tidal tail runs diagonally from SE to NW over $\sim4^\circ$,
between about ($RA=214^\circ$, $Dec=27^\circ$) and ($RA=210^\circ$,
$Dec=30^\circ$) (Belokurov et al.\ 2006). No H\,{\sc i} emission was detected
either within our outside this. Although we could search even larger areas, it
would be hard to establish a link between cluster and gas for separations more
than a few degrees.

%
% FIGURE 5
%
\begin{figure}
\centerline{\psfig{figure=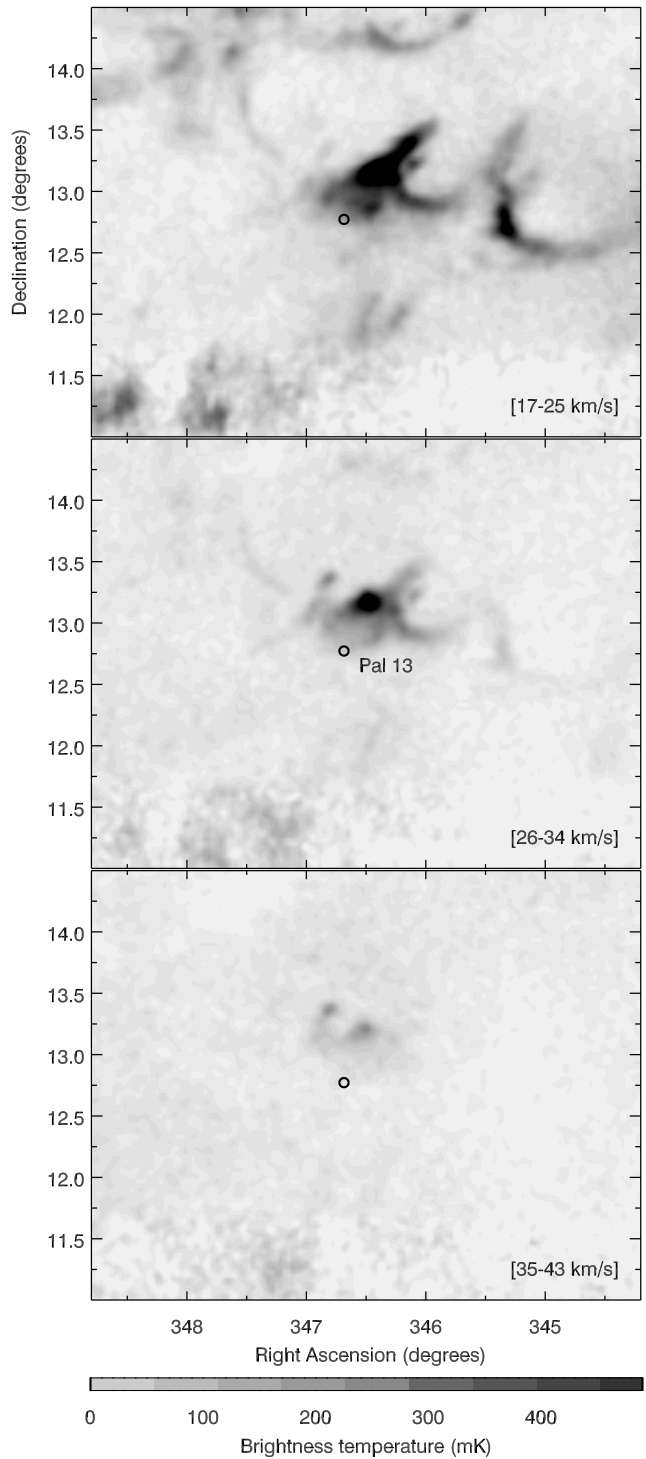,width=84mm}}
\caption[]{The 21-cm images around Pal\,13, in velocity ranges near H\,{\sc i}
emission at the systemic velocity of Pal\,13 (indicated with a circle;
middle), and slightly shifted in velocity (top/bottom).}
\end{figure}

The Pal\,13 data are confused with Galactic Disc emission. Although some
emission is concentrated near the cluster in position and velocity, at
slightly smaller velocities this emission is clearly part of an extended
interstellar cirrus complex (Fig.\ 5), which is around $l=87^\circ$,
$b=-43^\circ$. This cluster is one of the least massive globular clusters
known, and is unlikely to have accumulated much GCG.

M\,3 and NGC\,5466 were observed at Arecibo before, by Birkinshaw et al.\
(1983) who derived upper limits of 45 M$_\odot$ within the tidal sphere and
0.61 M$_\odot$ per beam for M\,3, and 1190 and 1.7 M$_\odot$, respectively,
for NGC\,5466. The new scanning observations place more stringent limits on
the mass contained within the large tidal radii. Birkinshaw et al.\ also
observed Pal\,4, but they searched a velocity range $v_{\rm LSR}>105$ km
s$^{-1}$ which excludes that of Pal\,4. Thus, ours is the first meaningful
upper limit on H\,{\sc i} emission from Pal\,4. We observed Pal\,13 at Arecibo
before (van Loon et al.\ 2006), failing to detect anything more significant
than a 3-$\sigma$ level of 1 M$_\odot$ (we did detect the Northern tip of the
Magellanic Stream, behind the cluster). Our new upper limits for Pal\,13 are
higher, but more realistic given the complex contamination from interstellar
cirrus.

%=========================================================================== 6
\section{Discussion: H\,{\sc i} in remote clusters}

No H\,{\sc i} emission could be detected in the massive cluster M\,3, and
searches for H\,{\sc i} emission from other massive clusters have been equally
unsuccessful (save for M\,15). A typical globular cluster produces
$\sim10^{2-3}$ M$_\odot$ of gas between successive cleansings when passing
through the Galactic plane (Tayler \& Wood 1975; McDonald et al.\ 2009). There
is evidence for clusters to be stripped of their gas on much shorter
timescales, possibly through interaction with the Galactic Halo (van Loon et
al.\ 2006; Boyer et al.\ 2006). Our non-detection in M\,3 once again confirms
the rapid removal timescale. Pal\,4 on the other hand is not very massive:
$M\sim2-3\times10^4$ M$_\odot$, from the relation with $M_{\rm V}$ (Mandushev
et al.\ 1991). Our measurements allow $\sim40$ M$_\odot$ of atomic hydrogen to
be present within the tidal volume of Pal\,4, i.e.\ of similar order to what
is expected to have accumulated over a Gyr.

Perhaps the presumptions about the most likely hosts of significant amounts of
GCG have been wrong. The nearer clusters experience much higher degrees of
harrassment by the Galactic Halo and Disc than very distant clusters. Even the
high metallicity of some prime targets may have fooled us in the past, if the
winds of metal-poor stars are slower (Marshall et al.\ 2004) and hence more
likely to be retained within the cluster's gravitational potential (cf.\
McDonald \& van Loon 2007) --- explaining the detection in one of the most
metal-poor clusters, M\,15 (van Loon et al.\ 2006).

Below, we discuss the prospects of finding gas associated with distant
globular clusters, and in particular the likelihood of association of Pal\,4
with the HVC next to it.

%------------------------------------------------------------------------- 6.1
\subsection{Constraints on the distance to the HVC and the likelihood of its
association with Pal\,4}

HVCs were found within $\sim50$ kpc from M\,31 (Westmeier, Br\"uns \& Kerp
2008), and where distances have been derived to Galactic HVCs they tend to be
$\sim10$ kpc (Putman et al.\ 2003; Thom et al.\ 2006, 2008; Wakker et al.\
2007, 2008). Either the HVC we discuss here is associated with Pal\,4 (which
would explain that it is not as nearby as other HVCs) or it is much closer to
us like other typical HVCs (and thus not associated with Pal\,4). Can we
distinguish between these two possibilities?

Pal\,4 is seen in a direction away from the Galactic Centre and Galactic plane
($l=202^\circ$ and $b=+72^\circ$). An extended complex of extra-planar gas is
seen in that general direction of the sky, e.g., in the Leiden-Argentine-Bonn
survey (Kalberla et al.\ 2005). The HVC near Pal\,4 is marginally consistent
with the ``population P'' proposed by Wakker \& van Woerden (1991), which they
note tend to be small clouds and with velocities suggesting they may belong to
the outer realms of the Galaxy. De Heij et al.\ (2002a) also found several
HVCs and compact HVCs with $90\ \lsim\ v_{\rm LSR}\ \lsim\ 130$ km s$^{-1}$ at
distances of $\geq 5^\circ$ from Pal\,4, which they argue are consistent with
a population of dark-matter halos in the Local Group (de Heij, Braun \& Burton
2002b).

As noted in the previous section, the HVC near Pal\,4 must be moving away from
the Milky Way at high speed. This is unlikely for the majority of the HVCs,
which Braun \& Burton (1999) point out may be the reason for the paucity of
HVCs observed in the direction of the Galactic poles. As Pal\,4 is also moving
away from the Milky Way a link is plausible. The orbital velocity of Pal\,4 is
not known but likely to be of order 200 km s$^{-1}$, and it must therefore
have a significant (but hard to measure) transversal velocity. If this is true
also for the HVC, then one could expect significant deformation due to the
interaction (through ram-pressure) with the Halo gas even at the relatively
low densities at the distance of Pal\,4. Indeed, the HVC resembles the
head-tail morphology seen in many HVCs and explained in this way (Westmeier et
al.\ 2005). From this, we would predict a motion of the HVC in a South-Eastern
direction.

%....................................................................... 6.1.1
\subsubsection{Constraints from cloud dynamics and morphology}

\underline{Pressure equilibrium:} The atomic density in the HVC, for a volume
of $\sim0.1\ (d/109\ {\rm kpc})^3$ kpc$^3$, is $n_{\rm
Cloud}\sim3\times10^{-2}\ (109\ {\rm kpc}/d)$ cm$^{-3}$. For it to be in
pressure equilibrium with the surrounding Halo gas, the Halo density must be
$n_{\rm Halo}=n_{\rm Cloud}\times\left(T_{\rm Cloud}/T_{\rm
Halo}\right)\sim3\times10^{-4}\ (109\ {\rm kpc}/d)$ cm$^{-3}$. This is $\sim5$
times higher than other estimates of the Halo density (Bregman 2007; cf.\
Sembach et al.\ 2003). The density profile of the Halo falls approximately as
$1/d$ (see Sternberg, McKee \& Wolfire 2002), so placing the HVC nearer or
farther away will not improve the agreement.

%
% FIGURE 6
%
\begin{figure}
\centerline{\psfig{figure=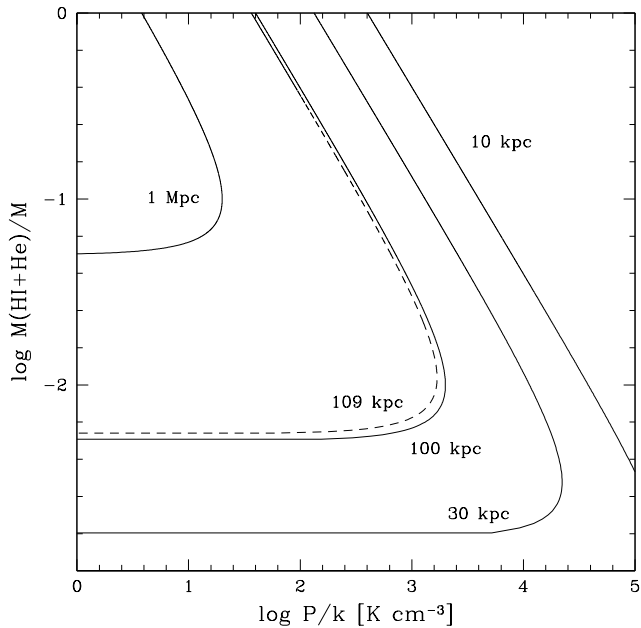,width=84mm}}
\caption[]{Diagram computed for the HVC near Pal\,4, of the external pressure,
$P$, required to keep the cloud stable, for different fractions of H\,{\sc i}
plus helium mass compared to the total mass, and for different distances.
Pal\,4 is at 109 kpc (dashed curve). This assumes the virial theorem holds,
and that $\sigma_{\rm v}=10$ km s$^{-1}$.}
\end{figure}

Gravitation helps to keep the cloud together. Following Westmeier et al.\
(2005) we computed the external pressure required to keep the HVC bound in
addition to gravitation by the cloud mass, assuming the virial theorem holds
(and setting $\sigma_{\rm v}\equiv10$ km s$^{-1}$). This depends on the
distance and on the fraction, $f$, of H\,{\sc i} plus helium mass compared to
the total mass (Fig.\ 6). For $\log f\gsim-1.5$ we obtained reasonable values
for the Halo pressure for distances as short as 10 kpc and as long as several
100 kpc (see Sternberg et al.\ 2002, their figure 14).

\underline{Thermal and dynamical instabilities:} One may expect thermal
instabilities to occur due to cooling, as well as Kelvin-Helmholtz
instabilities due to friction with the hot Halo. The thermal spatial and
temporal scales scale with $n_{\rm Cloud}^{-1}$, whilst the Kelvin-Helmholtz
timescale scales with the square root of the cloud--halo contrast (cf.\
Stanimirovi\'c et al.\ 2008). Both are $\sim10^8$ yr for $\sim$ kpc fragments,
for the density derived at $d=109$ kpc and a (distance independent)
cloud--halo density contrast $\sim100$. This could explain why the cloud is
confined to $\sim1$ kpc size, and the timescale is similar to the plausible
duration for the cloud to have been separated from Pal\,4.

Fluctuations in brightness temperature are seen across the surface of the HVC
on smaller scales, about 6--20$^\prime$. If these are the relevant scales at
which thermal instabilities are operating in the HVC, they would correspond to
physical sizes $\sim200$--600 $(d/109$ kpc) pc. In fact, Stanimirovi\'c et
al.\ (2008) estimate scales of $\sim100$--200 $(5\times10^{-2}$ cm$^{-3}/n)$
pc. These two estimates can be reconciled with the above estimated density of
the HVC, suggesting thermal instabilities on scales of 200--300 $(d/109$ kpc)
pc. Interestingly, this does not depend on distance, but it does require that
the majority of the gas is accounted for by the warm neutral medium traced by
H\,{\sc i}.

\underline{Differential drag:} The arc-like feature (``tail'') to the North of
the bulk of the HVC emission, at the side of Pal\,4, might be interpreted as
the result of differential drag as the cloud moves through the Halo. Peek et
al.\ (2007) presented a simple prescription of what might be expected:
\begin{equation}
n_{\rm H}\ d = \frac{1}{C_{\rm D}} \left(\frac{\Delta v}{v}\right)^2
\frac{\sin \phi}{\Theta \Delta N_{\rm H}^{-1}},
\end{equation}
where $C_{\rm D}$ is a drag coefficient, $\phi$ is the angle between the
cloud's motion and the line-of-sight, and $\Theta$ is the angular separation
between the HVC and its tail.

We estimate that the relative velocity difference is $\Delta v/v \simeq
10/100$, $\Theta \simeq 0.5^\circ$, and the neutral hydrogen column densities
$N_{\rm H} \sim 3\times10^{18}$ and $\sim10^{19}$ cm$^{-2}$ in the tail and
main body, respectively. Assuming $\sin \phi \sim0.5$ and $C_{\rm D}\simeq 1$
(see Peek et al.\ 2007), we thus obtain $n_{\rm H}\ d\sim 3\times10^{18}$
cm$^{-2}$ or $n_{\rm H}\sim1\times10^{-5}\ (109\ {\rm kpc}/d)$ cm$^{-3}$. This
is low, but given the uncertainties it is still consistent with measured Halo
gas densities. For instance, $C_{\rm D}$ might be a few times smaller and
hence the derived density a few times higher. The $1/d$ dependence of $n_{\rm
H}$ renders $n_{\rm H}\ d$ distance independent.

\underline{Multiphase medium:} Clouds near the Galactic plane can (but do not
always) exhibit multi-phase structure (Wolfire et al.\ 1995). Recent studies
suggest a multi-phase medium may exist even at distances as large as 150 kpc
(Sternberg et al.\ 2002). The HVC near Pal\,4 seems uniformly warm, but the
presence of a cold phase can not be excluded. The apparent warmth could be
entirely due to non-thermal motions (e.g., turbulence), although it would be
coincidental if it led to the observed temperature of 10,000 K so common for
warm interstellar gas. The presence of a warm neutral medium seems therefore
likely, but if this is the case then it could easily mask a modest amount of
cold medium. Either way, this lends us no clues with regard to its distance.

Relatively nearby clouds tend to shine in H$\alpha$ due to heating by the
Galactic Disc (radiatively) or Halo (kinetically) (Bland-Hawthorn et al.\
1998; Tufte et al.\ 2002; Putman et al.\ 2003). In the Magellanic Stream,
shock cascades running downstream cause it to light up in H$\alpha$
(Bland-Hawthorn et al.\ 2007). One would not expect an isolated cloud at 109
kpc to be bright in H$\alpha$, the meta-Galactic radiation field being too
weak. The expected brightness of the HVC is $S=n^2 L$ mR, where the density,
$n$, is in cm$^{-3}$ and the extend along the line-of-sight, $L$, is in kpc.
Assuming a spherical cloud, we obtain $S=10\ (109\ {\rm kpc}/d)$ mR. This is
faint, but a little above the detection limit of the most sensitive all-sky
H$\alpha$ survey, viz.\ the Wisconsin H$\alpha$ Mapper (WHAM; Haffner et al.\
2003). Unfortunately, the velocity coverage of WHAM was too restrictive, only
barely reaching the velocity of Pal\,4.

%....................................................................... 6.1.2
\subsubsection{Constraints from sodium absorption}

%
% FIGURE 7
%
\begin{figure}
\centerline{\psfig{figure=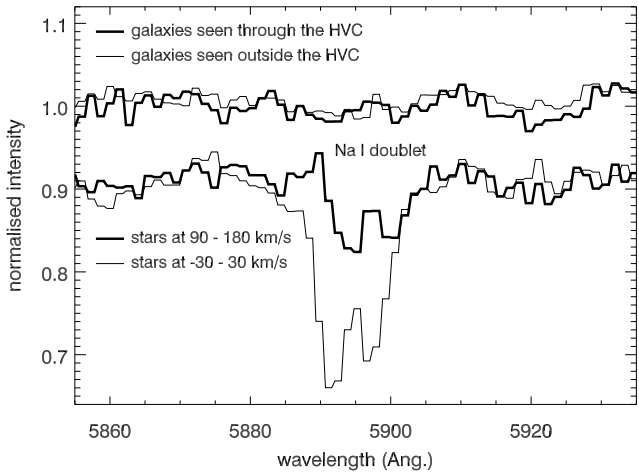,width=84mm}}
\caption[]{Sloan Digital Sky Survey (SDSS) spectra around the Na\,{\sc i} D
doublet, of the averages of galaxies seen through the HVC and near to it, and
of example averages of stars (offset by $-0.1$ for clarity) with velocities
around zero and around that of the HVC to illustrate the ability to detect the
redshift of the HVC. The line absorption in the stellar spectra arises in the
stellar photospheres; the galaxy spectra (a mixture of redshifts) show very
little interstellar absorption.}
\end{figure}

The presence or absence of absorption signatures of the cloud in the spectra
of background objects at known distances can help constrain the distance to
the cloud (e.g., Thom et al.\ 2006, 2008; Wakker et al.\ 2008). For this
experiment we considered Sloan Digital Sky Survey Data Release 7 (SDSS-DR7;
Abazajian et al.\ 2009) spectra of the strong Na\,{\sc i} D doublet around 589
nm that is commonly seen in the neutral and weakly-ionized medium.

The spectral resolution is $\sim30$ km s$^{-1}$; this is about three times the
expected intrinsic line width judged from the observed velocity dispersion of
the HVC on arcminute scales, but the intrinsic line width will be narrower
still if $\sigma_{\rm v}$ drops at sub-arcminute scales and/or if $\sigma_{\rm
v}$ is dominated by thermal motions in which case the heavier sodium atoms
display smaller velocities than the hydrogen atoms.

We first compiled a composite of 16 stars of spectral types A--G, with
velocities between $-30\leq v_{\rm LSR}\leq 30$ km s$^{-1}$, to show a typical
Na\,{\sc i} D doublet (Fig.\ 7). In this case, the absorption arises mostly or
entirely in the photosphere of the stars; to test our ability to discern
redshifted absorption, we also compiled a composite of similar stars with
velocities between $90\leq v_{\rm LSR}\leq 180$ km s$^{-1}$ and indeed the
photospheric absorption is redshifted by a detectable amount.

We then retrieved the spectra of 298 {\it galaxies}, that are definitely
behind the HVC, to check whether any intervening absorption could be detected
at all. We only selected galaxies with redshifts that avoided direct blends
with strong spectral lines in the intrinsic galaxy spectrum. Otherwise, the
redshifts vary and the expectation was that this would lead to annihilation of
the features in the galaxy composite whilst maintaining any intervening
absorption which does not partake in the galaxy's redshift. This worked very
well (Fig.\ 7): the galaxy composite shows a featureless continuum to within
approximately $\pm2$\%.

Comparing galaxies seen through the HVC (within $0.3^\circ$ from its centre)
and those surrounding it, we see that the difference is very small and one
cannot say with any certainty whether there is absorption redshifted by the
amount expected for the HVC. It is clear that the experiment using Halo stars
would not give any useful result. We thus place an upper limit to the Na\,{\sc
i} D absorption of $W\ \lsim\ 0.1$ \AA\ or $N_{\rm NaI}\ \lsim\ 10^{12}$
cm$^{-2}$.

For scaled-solar abundances $N_{\rm Na}\sim2\times10^{-5}\ N_{\rm H}\ (Z/{\rm
Z_\odot})$ (Grevesse \& Noels 1993). With $N_{\rm H}\sim10^{18-19}$ cm$^{-2}$
in the HVC near Pal\,4 this would imply $Z_{\rm HVC}<0.1$ Z$_\odot$. HVCs in
the inner Halo generally have metallicities 0.1--1 Z$_\odot$ (Richter et al.\
2009, and references therein), but Pal\,4 has a metallicity $Z_{\rm
Pal\,4}=0.03$ Z$_\odot$. The non-detection of Na\,{\sc i} in the HVC is thus
more consistent with an origin in Pal\,4 than with typical HVCs closer to the
Milky Way. But the uncertainties are substantial and the upper limit on the
sodium column density is not inconsistent with typical values found in
extra-planar gas (Ben Bekhti et al.\ 2008).

%....................................................................... 6.1.3
\subsubsection{Summary on constraint on the HVC distance}

We have considered several possibilities to constrain the HVC distance. The
assumption of pressure equilibrium, the spatial and temporal scales at which
instabilities occur, the likelihood of a multi-phase medium, and limits on
the metal content all point at a distance $10\ \lsim\ d\ \lsim\ 200$ kpc. This
does not discriminate between an association of the HVC with Pal\,4 and a
chance alignment of the two.

%------------------------------------------------------------------------- 6.2
\subsection{Distant globular clusters as probes of the outer Galactic Halo}

The most distance cluster in our sample, Pal\,4 is the only one with a
possible gas association. Distant globular clusters might well be the most
likely places to find large amounts of GCG, because they have not passed
through the Galactic Disc for some time, and the Galactic Halo is relatively
tenuous that far from the Galactic Centre. The same is true for other
satellites of the Milky Way, such as low-mass dwarf galaxies some of which
have been found at distances comparable to the Magellanic Clouds (e.g.,
Willman et al.\ 2005).

Pal\,4, like the other distant clusters Pal\,3, Eridanus, Pal\,14, and AM\,1,
appears to be 1.5--2 Gyr younger than the inner Halo globular clusters
(Forbes, Strader \& Brodie 2004; Dotter, Sarajedini \& Yang 2008). Red giants
in these clusters would lose more mass than their older, lower-mass siblings,
thus enhancing the rate of GCG replenishment. This could alleviate the problem
of the large inferred mass for the HVC, but only slightly.

Also, Pal\,4 presently traverses the far outskirts of the Galactic Halo; it
will definitely have had much longer than usual to accumulate the slow ejecta
from its red giants. In fact, its trajectory is undetermined, and it may not
even be in orbit around the Milky Way, and never have passed through the
Galactic Disc.

In addition, the HVC may have collected much of its mass from Halo gas swept
up by Pal\,4: e.g., for a column of 0.2 kpc diameter (Pal\,4's tidal
cross-section) and 1.3 Mpc length (two circular orbits), it would have swept
up a similar amount of matter as is currently observed. This requires that the
gas is collected through ram-pressure interaction with gas already filling the
tidal sphere; accretion by gravity would not work as the speed of Pal\,4
relative to the Halo gas exceeds the cluster's escape velocity by an order of
magnitude and gas entering an ``empty'' Pal\,4 would just as easily leave it
again through the backdoor.

The displacement of $\sim2$ kpc between the HVC and Pal\,4 (if at a common
distance) corresponds to a timescale of $5\times10^7$ yr, assuming a relative
velocity of 40 km s$^{-1}$. This becomes longer if they differ in distance to
us and their relative motion across the sky is slower than along the
line-of-sight --- or potentially shorter if the relative motion across the sky
is faster. In fact, much of the observed radial velocity component is away
from the Galactic Centre, meaning that the HVC is moving away from the Milky
Way at considerable speed, $\sim90$ km s$^{-1}$ (if it is at $d\ \gsim\ 100$
kpc), and faster than Pal\,4 does ($\sim50$ km s$^{-1}$). This implies that if
the HVC was once part of Pal\,4, that it must have been accelerated during or
since it had become detached from the cluster.

We speculate that possible acceleration mechanisms might include large-scale
convection (from a combination of buoyant hot gas cells and cooling flows)
within the Halo, or a coronal (pressure-driven) or disc (radiation-driven)
Galactic wind (cf.\ Veilleux, Cecil \& Bland-Hawthorn 2005). Either of these
scenarios would imply an external pressure (by gas or radiation) being the
source of acceleration, rather than an internal source (as in a rocket). The
observed head-tail morphology of the HVC would arise naturally: assuming the
HVC is in the background of Pal\,4, then its "head" points roughly at the
Milky Way (and Pal\,4).

GCG removal timescales of the order of $10^6$ yr have been inferred for
clusters much deeper in the Galactic Halo, typically $\sim10$ kpc from the
Milky Way (van Loon et al.\ 2006; Boyer et al.\ 2006; McDonald et al.\ 2009).
At the distance of Pal\,4, the Halo density is two orders of magnitude lower
(Sternberg et al.\ 2002); stripping is thus much less effective. It will take
gas in Pal\,4 two orders of magnitude longer to have run into the equivalent
amount of Halo material as nearby globular clusters, i.e.\ $10^8$ yr. This
compares well with the above estimate of the dynamical timescale if the HVC
has been removed from Pal\,4.

If the Halo gas is effective in removing GCG from globular clusters, then the
more likely place to catch it is not inside the cluster but at some distance
trailing it. This might explain the small offset from the cluster centre of
the cloud discovered in M\,15 (Boyer et al.\ 2006), as well as the cloud we
now discovered near Pal\,4. This would be slightly at odds with the
interpretation of X-ray emission {\it in front} of several clusters to be due
to bow-shock interaction between the GCG and the Halo gas (Okada et al.\
2007). Perhaps, whether a bow-shock or trailing cloud is observed depends on
the pressure balance between the GCG and Halo.

%....................................................................... 6.2.1
\subsubsection{Could we have discovered a dark galaxy?}

Dark mini-halos are predicted by $\Lambda$CDM cosmologies to swarm the Local
Group (Kauffmann, White \& Guiderdoni 1993). Some of these may never have
ignited to become one of the luminous dwarf galaxies that are being discovered
in increasing numbers (cf.\ Simon \& Geha 2007). Completely dark halos would
obviously be hard to find. Does the large mass for the HVC near Pal\,4 as
inferred from its apparent rotation bear evidence of a dark galaxy host to the
HVC and Pal\,4? If the relative velocity of Pal\,4 also reflects an orbital
motion around the cloud, then the enclosed mass may be as high as $10^8$
M$_\odot$ (or higher depending on the orientation of the orbit), i.e.\ at
least several hundred times the combined masses seen in the HVC's H\,{\sc i}
and Pal\,4's starlight. If this is the case, one must consider this system a
dark halo.

%
% FIGURE 8
%
\begin{figure}
\centerline{\vbox{
\psfig{figure=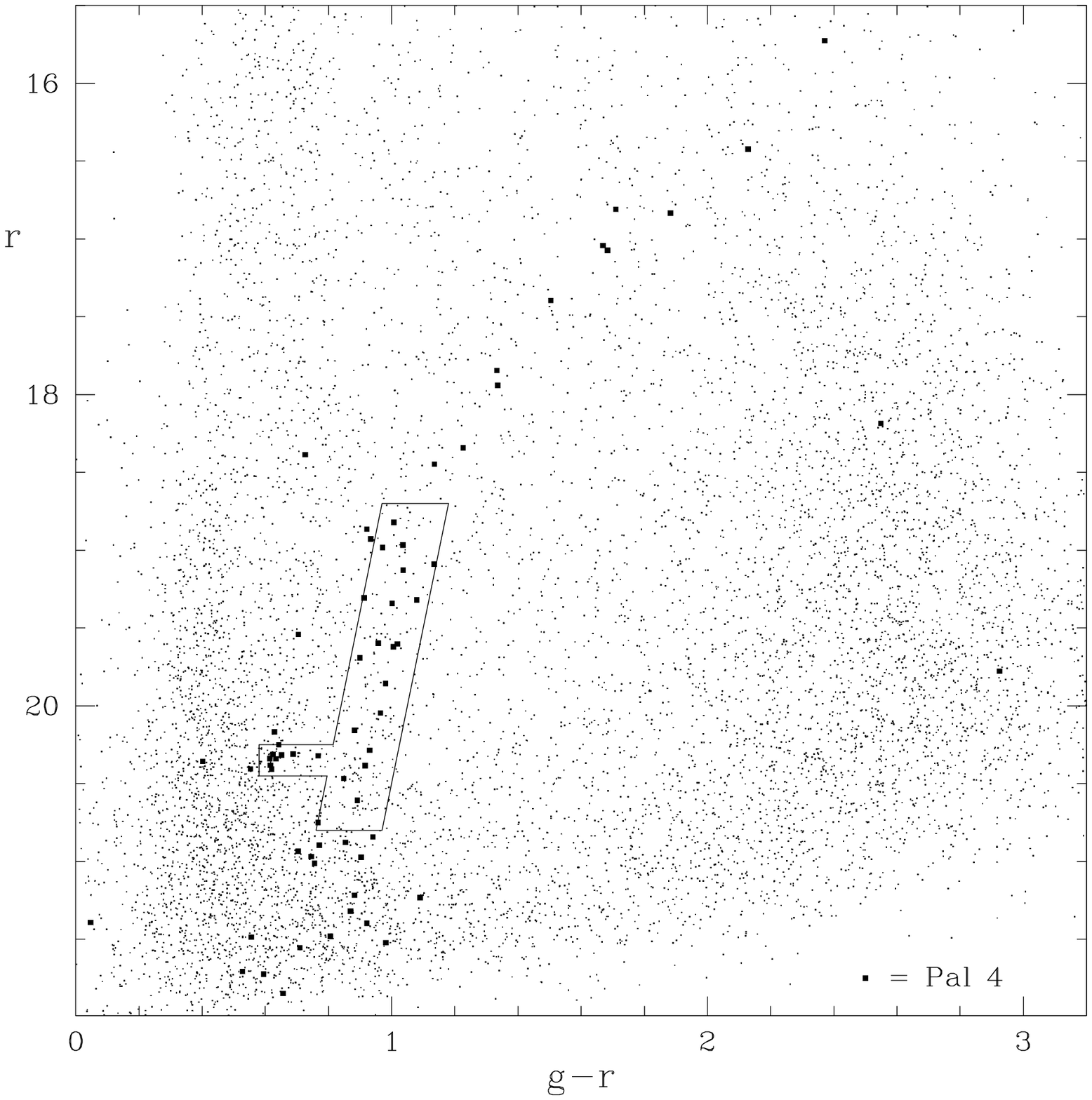,width=84mm}
\psfig{figure=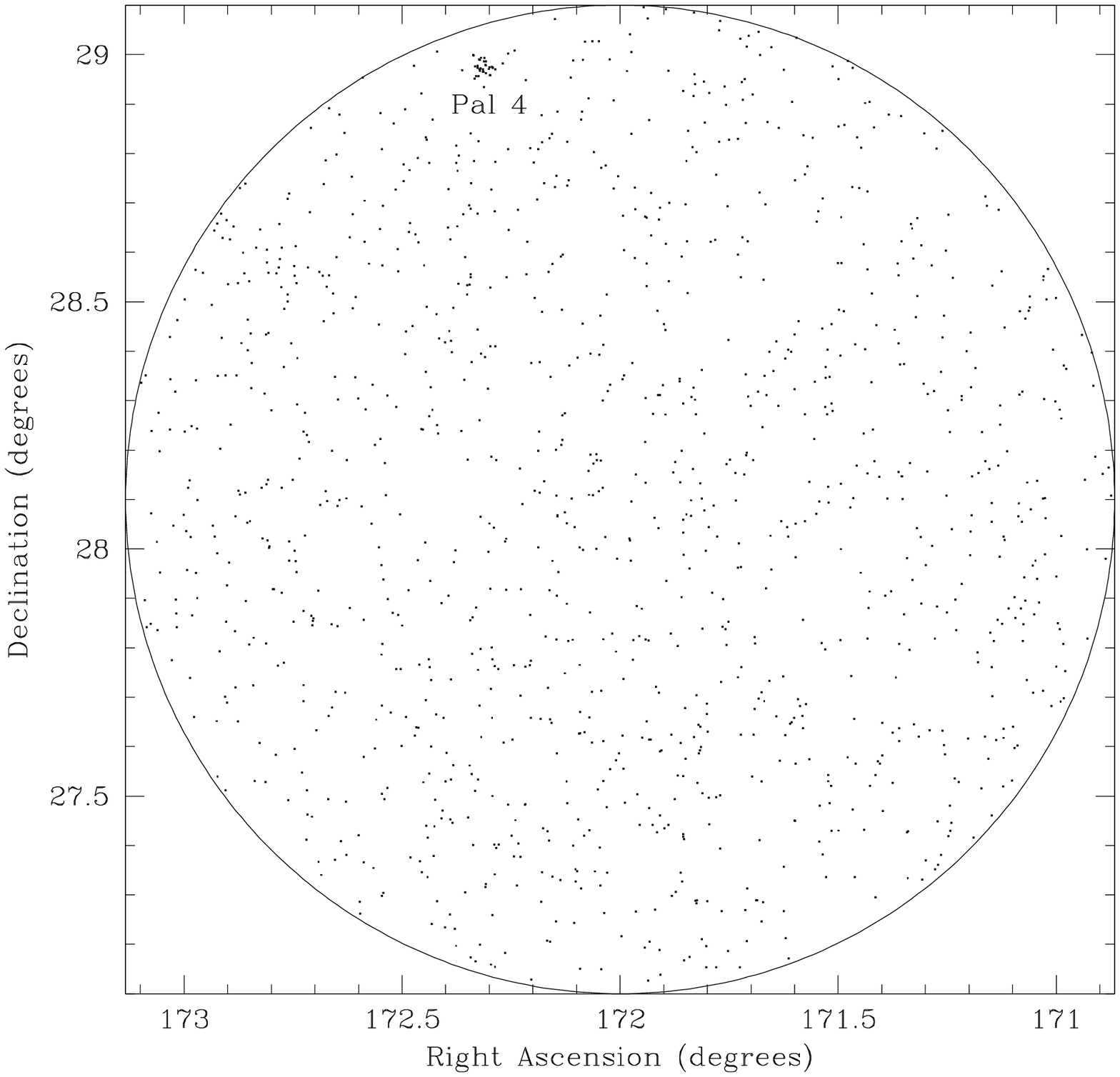,width=84mm}
}}
\caption[]{Sloan Digital Sky Survey (SDSS) photometry in Pal\,4, and within a
$1^\circ$ radius field centred on (RA,Dec)=(172$^\circ$,28.1$^\circ$) (top
panel). The positions on the sky of stars within the box are plotted in the
lower panel, comfortably encompassing the HVC.}
\end{figure}

The object may not be strictly dark. To search for a stellar counterpart to
the HVC, we have mined the Sloan Digital Sky Survey Data Release 6 (SDSS-DR6).
Based on the location in the $(g-r),r$ colour-magnitude diagram of stars
within $1.8^\prime$ from Pal\,4, we defined a locus of relatively numerous red
giants and horizontal branch stars (Fig.\ 8, top). Plotting stars that
satisfied this photometric criterion within a $1^\circ$ radius from
(RA,Dec)=(172$^\circ$,28.1$^\circ$) revealed no obvious overdensity apart from
Pal\,4 (Fig.\ 8, bottom). Some clustering may be present near
(RA,Dec)=(171.8$^\circ$,27.6$^\circ$), but this is hardly convincing and in
any case would not amount to more than the equivalent of Pal\,4.

The faint dwarfs that have recently been discovered near the Milky Way have
typically a radius of $\lsim10^\prime$, and a surface brightness $\mu_{\rm
V}\sim29$ mag arcsec$^{-2}$ but perhaps as dim as 31 mag arcsec$^{-2}$ (Liu et
al.\ 2008). If the light of Pal\,4 were to be spread out over a square degree,
it would have $\mu_{\rm V}\sim32.7$ mag arcsec$^{-2}$. A low-luminosity system
such as Willman\,1 (Willman et al.\ 2005) --- which incidentally has a
metallicity similar to that of Pal\,4 (Siegel, Shetrone \& Irwin 2008) ---
would have been very hard to detect indeed if it were as diffuse as the HVC.
It seems thus likely that, if the HVC near Pal\,4 contains $>10^6$ M$_\odot$,
it must be pretty dark.

%=========================================================================== 7
\section{Conclusions}

Using GALFA-H\,{\sc i} Survey data, we searched for neutral hydrogen gas
associated with four Galactic globular clusters, M\,3, NGC\,5466, Pal\,4, and
Pal\,13. No gas was found within the tidal radii of any of these clusters,
down to 3-$\sigma$ limits of 6--51 M$_\odot$. These represent the most
stringent limits on the gas content of these clusters to date. In the case of
the massive cluster M\,3 this confirms once again that efficient removal
mechanisms must be at work that clean globular clusters of gas on timescales
of $\ll 10^8$ yr.

We did discover a compact H\,{\sc i} high-velocity cloud (HVC) near Pal\,4,
which at 109 kpc distance from the Sun is one of the most remote clusters in
the Galactic Halo. The properties of the cloud resemble those of other HVCs
located nearer to the Milky Way, but its high velocity directed away from the
Galaxy and its relative isolation but proximity to Pal\,4 are tantalizing ---
a physical association between the two remains possible. However, one would
need to invoke rather extreme descriptions for the interaction of the
Pal\,4/HVC system with the Milky Way halo, to account for the following
observations:\\
$\bullet$ The HVC lies beyond the gravitational influence of Pal\,4;\\
$\bullet$ The H\,{\sc i} mass of the HVC exceeds that of Pal\,4;\\
$\bullet$ Their galacto-centric velocities differ by 40 km s$^{-1}$.

The internal kinematics of the cloud suggests an even higher mass,
$3\times10^6$ M$_\odot$ if at the distance of Pal\,4. Most of this mass must
be either cold ($\lsim 100$ K), very hot ($\gsim\ 10^6$ K), or dark
(non-baryonic). If it is cold it might be possible to detect in a high angular
resolution H\,{\sc i} absorption-line experiment. If it is very hot it might
be detectable in X-rays --- although there is no {\it a priori} reason to
expect such hot plasma. On the other hand, if the H\,{\sc i} structure seen in
the HVC on 0.1--$1^\circ$ scales is due to thermal instabilities then most of
the gas is likely to be traced by H\,{\sc i}. But this does not preclude the
cloud and Pal\,4 being the flimsy tracers of a dark mini-halo, a relic of the
re-ionization epoch in which star formation might only have led to the
production of Pal\,4\footnote{We would like to point out that Pal\,4 is in the
constellation of Ursa Major, but the H\,{\sc i} column density of the HVC has
its maximum just across the border, in the constellation of Leo.}. This may
fit a scenario proposed recently by Ricotti (2009) for the late gas accretion
by primordial mini-halos to explain faint dwarf galaxies like Leo\,T and some
HVCs. In this scenario, Pal\,4 would orbit the dark mini-halo, the inferred
mass of which would then be quite phenomenal ($\gsim10^8$ M$_\odot$). This
``mini''-halo would indeed be largely dark, having accreted only the gas we
see in the form of the HVC.

%=============================================================================
\section*{Acknowledgments}

It is a great pleasure to thank Joana Oliveira for her help at various stages.
We also wish to thank the referee for her/his constructive report.
The Turn On GALFA Survey (TOGS) H\,{\sc i} data are part of the Galactic ALFA
(GALFA) survey data set obtained with the Arecibo L-band Feed Array (ALFA) on
the Arecibo 305m telescope. Arecibo Observatory is part of the National
Astronomy and Ionosphere Center, which is operated by Cornell University under
Cooperative Agreement with the National Science Foundation of the United
States of America.
Funding for the SDSS and SDSS-II has been provided by the Alfred P.\ Sloan
Foundation, the Participating Institutions, the National Science Foundation,
the U.S.\ Department of Energy, the National Aeronautics and Space
Administration, the Japanese Monbukagakusho, the Max Planck Society, and the
Higher Education Funding Council for England. The SDSS Web Site is
http://www.sdss.org/.

%=============================================================================

\label{lastpage}

\end{document}